\documentclass[]{raa}   
\usepackage{graphicx,times}             %for PS/EPS graphics inclusion, new
\input{epsf.sty}                        %for PS/EPS graphics inclusion, old
\input{psfig.sty}                       %for PS/EPS graphics inclusion, old

\begin{document}

\title{A two-component jet model based on the Blandford-Znajek and Blandford-Payne processes
%\footnotetext{$*$ Supported by the National Natural Science Foundation of China.}
}
%   \subtitle{I. Place Your Subtitle Here}

   \volnopage{Vol.0 (200x) No.0, 000--000}      %%preserved for Editor. DOn't remove!
   \setcounter{page}{1}          %%starting page, preserved for Editor. DOn't remove!

   \author{Wei Xie
      \inst{1}
   \and Wei-Hua Lei
      \inst{1,2}
   \and Yuan-Chuan Zou
      \inst{1}
   \and Ding-Xiong Wang
      \inst{1}
   \and Qingwen Wu
      \inst{1}
   \and Jiu-Zhou Wang
      \inst{1}
   }

\institute{School of physics, Huazhong University of Science and Technology,
  Wuhan 430074, China; {\it leiwh@hust.edu.cn}\\
        \and
             Department of Physics and Astronomy, University of Nevada Las Vegas, 4505 Maryland Parkway, Box 454002, Las Vegas, NV89154-4002, USA.\\
   }

   \date{Received~~2009 month day; accepted~~2009~~month day}

\abstract{ We propose a two-component jet model consistent with the observations of several gamma ray bursts (GRBs) and active galactic nuclei (AGNs). The jet consists of inner and outer components, and they are supposed to be driven by the Blandford-Znajek (BZ) and Blandford-Payne (BP) processes, respectively. The baryons in the BP jet is accelerated centrifugally via the magnetic field anchored in the accretion disk. The BZ jet is assumed to be entrained a fraction of accreting matter leaving the inner edge of the accretion disk, and the baryons are accelerated in the conversion from electromagnetic energy to the kinetic energy. By fitting the Lorentz factors of some GRBs (GRB 030329, GRB 051221A, GRB 080413B) and AGNs (Cen A, Mkn 501 and Mkn 421) with this model, we constrain the physical parameters related to the accretion and outflow of these two kind of objects. We conclude that the spine/sheath structure of the jet from these sources can be interpreted naturally by the BZ and BP processes.
}

   \authorrunning{W. Xie, W.-H. Lei, Y.-C. Zou, D.-X. Wang, Q. Wu \& J.-Z. Wang }            %author_head in even pages
   \titlerunning{A two-component jet model based on the Blandford-Znajek and Blandford-Payne processes }  % title_head in odd pages

   \maketitle
%% The author head (on even pages) and the title head (on odd pages) will be
%% automatically extracted from \author{} and \title{}. Whenever the title is too long,
%% you will be asked to supply a shorter one by inserting either \authorrunning{} or
%% \titlerunning{} before \maketitle. Anyway, you can specify your own heads.
%%
%%
%% Note: In the following text body of your manuscript, please note several differences from
%%       other major journals:
%% (1) \subsection{Please Capitalize the First Letter of Each Notional Word in Subsection Title}
%% (2) Please Capitalize the First Letter of Each Notional Word in all tables' captions

%
%________________________________________________ sections below
%
\section{Introduction}           %% first-level sections will be auto-capitalized
\label{sect:intro}

Jets (outflows) exist in a variety of astrophysical objects in different sizes such as Active Galactic Nuclei (AGNs), Gamma Ray Bursts (GRBs), X-Ray Binaries (XRBs), Young Stellar Objects (YSOs), and so on. Although in most cases jets are assumed to be homogeneous conical outflows, in reality they can be structured (Zhang, Woosley, \& MacFadyen \cite{Zhang03}; Zhang, Woosley, \& Heger \cite{Zhang04}). It is usually assumed that the energy per unit solid angle depends as a power-law or a Gaussian function on the angular distance from the axis (M\'{e}sz\'{a}ros et al. \cite{Meszaros98}; Dai \& Gou \cite{Dai01}; Rossi, Lazzati \& Rees \cite{Rossi02}; Zhang \& M\'{e}sz\'{a}ros \cite{ZhangB02};
Kumar \& Granot \cite{Kumar03}; Salmonson \cite{Salmonson03}; Granot \& Kumar \cite{Granot03}; Zhang et al. \cite{ZhangB04}). Meanwhile, as an alternative structured jet model, two-component jet has been referred often. Berger et al. \cite{Berger2003} proposed that the observations of GRB 030329 require a two-component explosion: a narrow ($5^\circ$) ultra-relativistic component responsible for the $\gamma$-rays and early afterglow, and a wide, mildly relativistic component responsible for the radio and optical afterglow beyond 1.5 days. A detailed calculation about relativistic two-component jet was proposed by Peng et al. (\cite{Peng05}). Huang et al. (\cite{Huang04}) consider the rebrightening of XRF 030723 as a further evidence for a two-component jet in a GRB: with a narrow but ultra-relativistic inner outflow and a wide but less energetic outer ejecta, a two-component jet will be observed as a typical gamma-ray burst if our line of sight is within the angular scope of the narrow outflow; otherwise, if the line of sight is within or slightly beyond the cone of the wide component, an X-ray flash will be detected. Wu et al. (\cite{WuXF05}) discussed the polarization of GRB afterglows from two-component jets. Racusin et al. (\cite{Racusin08}) claimed that the chromatic behavior of the broadband afterglow of GRB 080319B is consistent with viewing the GRB down the very narrow inner core of a two-component jet that is expanding into a wind-like environment. The broad-band light curve of the afterglow of GRB 080413B was well fitted with an on-axis two-component jet model (Filgas et al. \cite{Filgas11}).

Structured jets are also frequently referred to in AGNs. In order to reconcile the viability of the unification scheme of BL Lacs and FR I radio galaxies, Chiaberge et al. (\cite{CCCG00}) suggested a two-component jet model in which a fast spine is surrounded by a slow (but still relativistic) layer so that the emission at different angles is dominated by different velocity components: the fast one dominates the emission in BL Lacs while the slow layer dominates the emission in misaligned objects (FR I radio galaxies for example). According to the unification scenario, the BL Lacs and FR I radio galaxies are intrinsically the same, and the observation differences of these two objects just result from the different orientations of the observer. By means of modeling the observed spectral energy distribution (SED), people could derive the value of the jet Lorentz factor of the BL Lacs with typical value of $10\sim20$ (Hovatta et al. \cite{Hovatta09}). However, with the single emission component model, this Lorentz factor could not satisfy the observations of FR I galaxies which generally require relatively lower Lorentz factor (Xu et al. \cite{Xu00}). Consequently, a velocity structured jet model, as a simple hypothesis, could plausibly account for above discrepancy. The direct observational radio maps of the jet in several radio galaxies shown a limb-brightened morphology, which can be naturally interpreted as evidence of a slower external flow surrounding a faster spine (e.g., Giroletti et al. \cite{Giroletti04}). In addition, the structured model are also proposed to explain the high energy radiation (Ghisellini et al. \cite{Ghisellini05}; Hardcastle \cite{Hardcastle06}; Jester et al. \cite{Jester06}, \cite{Jester07}; Siemiginowska et al. \cite{Siemiginowska07}; Kataoka \cite{Kataoka08}). A succession of VLBI studies hinted that the pc-scale jets in strong TeV BL Lacs move slowly (Edwards \& Piner \cite{Edwards02}; Piner \& Edwards \cite{Piner04}; Giroletti et al. \cite{Giroletti04}). However, the bright and rapidly variable TeV emission indicates that at the jet scales where this emission originates, the jet should be highly relativistic (Dondi \& Ghisellini \cite{Dondi95}; Tavecchio et al. \cite{Tavecchio98}, \cite{Tavecchio01}; Kino et al. \cite{Kino02}; Ghisellini et al. \cite{Ghisellini02}; Katarzynski et al. \cite{Katarzynski03}; Krawczynski et al. \cite{Krawczynski02}; Konopelko et al. \cite{Konopelko03}). In view of the above observations, Georganopoulos \& Kazanas (\cite{GK03}) proposed a radially structured jet model in which the jet is rapidly decelerating in the $\gamma$-ray zone with a fast moving base. Ghisellini et al. (\cite{Ghisellini05}) argued that the jet could be structured in the transverse direction, being composed by a slow layer and a fast spine. For more applications of the two-component jet model, one can turn to the references in Chiaberge et al. (\cite{CCCG00}).

To sum up, the general picture of the two-component jet model can be described as: a narrow, highly relativistic jet surrounded by a wider, moderately relativistic outflow.

As mentioned above, the two-component jet model can successfully explain some observations in GRBs and AGNs. However, the physical origin of this structured jet has not been well understood yet. Sol et al. (\cite{Sol89}) proposed a two-flow model for extragalactic radio jets, in which one flow is a beam of relativistic particles coming out from the funnel or the innermost part of the accretion disc, the other flow is a classical or mildly relativistic disk wind coming out from all parts of the accretion disk. This work concluded that the beam-wind configuration is stable as long as the magnetic field (assumed longitudinal) is strong enough. However, they did not explain how the relativistic beam is formed. Eichler \& Levinson (\cite{Eichler99}) suggested a two-component jet model with a baryon-poor jet existing within a baryon-rich outflow. The baryon-poor jet may be driven by the Blandford-Znajek mechanism (Blandford \& Znajek \cite{BZ77}, hereafter BZ77; Macdonald \& Throne \cite{MT82}), in which the rotational energy of balck hole (hereafter BH) is extracted to power the jet in form of Poynting flux via the open field lines threading the horizon. Recently, Meier (\cite{Meier03}) discussed the probability of using the coexistence of Blandford-Znajek and Blandford-Payne (Blandford \& Payne \cite{BP82}, hereafter BP82) processes  as an interpretation of the two-component jets for quasars and microquasars. In the BP process, a baryon-rich outflow can be launched centrifugally via the open magnetic field threading the disk. It is argued that the baryon-rich jet can also play important role in the collimation of the central jet (Eichler \& Levinson \cite{Eichler99}; Tsinganos \cite{Tsinganos10}).
Motivated by the above works, we propose a two-component jet model for both GRBs and AGNs, in which the inner and outer jets are powered by the BZ and BP process respectively. Based on reasonable magnetic configuration and assumptions, we obtain the Lorentz factors for the inner and outer jets. By doing this, we can constrain the physical parameters of the central engine for GRBs and AGNs with the observations.

This paper is organized as follows. In Sect. 2 we describe the the two-component jet model in detail, and obtain the Lorentz factor of the outer-wide-slow (BP) and inner-narrow-fast (BZ) jets in Subsections 2.1 and 2.2, respectively. In addition, we compare the Lorentz factor of these two components and fit for several GRBs and AGNs in Sect. 3. Finally, the conclusions and discussions are presented in Sect. 4. Throughout this paper the units $G = c = 1$ are used.

%% Authors can give a citation as 'Michel et al. 1992'.
%% You may also use \cite, \citep and \citet for citation, and use Table~1 or Figure~1
%% and so forth. Using \ref and \label for cross-references of Tables/Figures
%% is a good way in adjusting/adding/removing text, tables or figures.

\section{The Two-Component Jet Model}
\label{sect:model}

%The photometry of three $\delta$ Sct stars AD Arietis, IP Virginis and
%YZ Bootis was performed from 2000 February 26 to
%2001 January 31\footnote{Please note the order: year, month, day}
%with the three photometers mounted on the 85-cm telescope at
%the Xinglong Station of BAO\footnote{Now NAOC}.
%The typical accuracy yielded from magnitude differences between
%reference stars is about 0.005\,mag.
%The observing log is given in $\cdots\cdots$

\begin{figure}
\centering
\includegraphics[width=80mm]{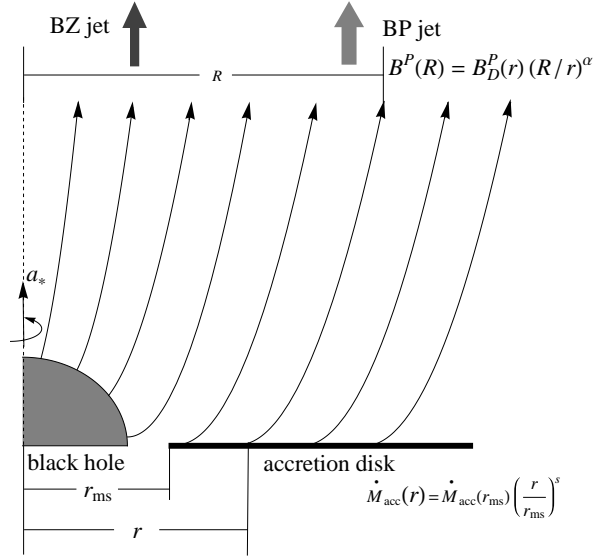}\\
    \caption{A schematic drawing of the magnetic field configuration for the two-component jet model, in which the inner-narrow-fast jet and the outer-wide-slow jet are driven by the BZ and BP processes, respectively.}
    \label{Fig:fig1}
\end{figure}

The schematic picture of the model is shown in Fig.~\ref{Fig:fig1}. The the BZ process launches the inner jet via the open magnetic field threading the BH, while the BP process produces the outer jet via the open magnetic field threading the disk. A similar magnetic configuration is also suggested by Li, Wang \& Gan. (2008) to study the jet power from AGNs.

\subsection{The Lorentz Factor of the outer jet driven by the BP process}
As argued by BP82, the baryons can be accelerated centrifugally along the magnetic field lines and form a magnetohydrodynamic (MHD) outflow, provided that the poloidal magnetic field is strong and inclined enough. To produce such a jet, the poloidal field lines are supposed to make an angle of less than $60^{\circ}$ to the outward radius vector at the disk mid-plane. It was argued by Cao (\cite{Cao97}) that this critical angle could be larger than $60^{\circ}$ for the rotating BH, implying that the flow can be easily accelerated in the BP process.

The configurations of the magnetic field are shown in Fig.~\ref{Fig:fig1}. Following BP82, we assume that the poloidal magnetic field on the disk surface varies with the disk radius as
\begin{equation}
B^{\rm{P}}_{\rm{D}}= B^{\rm{P}}_{\rm{H}}(r/r_{\rm{H}})^{-5/4},
\label{eq:1}
\end{equation}
\noindent where r is the disk radius and $r_{\rm H}=M(1+q)$ is the horizon radius of the Kerr black hole, here $M$ is the mass of the black hole, $q\equiv\sqrt{1-a_{*}^2}$, $a_{*}\equiv a/M$, $a=J/M$ is the angular momentum per unit mass of the black hole. The quantities $B_{\rm D}^{\rm P}$ and $B^{\rm{P}}_{\rm{H}}$ are the poloidal magnetic field at the disk and the black hole horizon, respectively.

The poloidal magnetic field far from the disk surface is assumed to be self-similar (BP82; Lubow et al. \cite{Lubow94}),
\begin{equation}
    B^{\rm{P}}= B^{\rm{P}}_{\rm{D}}(R/r)^{-\alpha},
\label{eq:2}
\end{equation}
\noindent where $\alpha$ ($\alpha \ge 1 $) is the self-similar index to describe the variation of the poloidal magnetic field with the cylindrical radius $R$ of the jet.

The magnetic field at BH horizon can be estimated by considering the balance between the magnetic pressure on the horizon and the ram pressure in the innermost parts of an accretion flow (Moderski et al. \cite{Moderski97})
\begin{equation}
    \frac{(B^{\rm{P}}_{\rm{H}})^2}{8\pi}=P_{\rm{ram}}\sim\rho \sim\frac{\dot{M}_{\rm{acc}}(r_{\rm{ms}})}{4\pi r_{\rm H}^2},
\label{eq:3}
\end{equation}
\noindent where $\dot{M}_{\rm{acc}}(r_{\rm{ms}})$ is the accretion rate at the inner edge of the disk, and $r_{\rm{ms}}$ is the radius of the innermost stable circular orbit (ISCO, Novikov \& Thorne \cite{NT73}, Bardeen et al. \cite{Bardeen72}), for prograde orbits, $r_{\rm ms}$  is given as $r_{ms}=M\{3+Z_2-[(3-Z_1)(3+Z_1+2Z_2)]^{1/2}\}$, here $Z_1\equiv1+(1-a_{*}^2)^{1/3}[(1+a_{*})^{1/3}+(1-a_{*})^{1/3}]$, and $Z_2\equiv(3a_*^2+Z_1^2)^{1/2}$.

Considering the mass outflow driven by BP process, we write the dependence of $\dot{M}_{\rm acc}(r)$ on radius as follows (Blandford \& Begelman \cite{Blandford99})
\begin{equation}
    \dot{M}_{\rm acc}(r)=\dot{M}_{\rm acc}(r_{\rm ms})\left(\frac{r}{r_{\rm ms}}\right)^s, \ \ 0 \le s \le 1.
\label{eq:4}
\end{equation}

According to the mass conservation law, the accretion rate of disk matter is related to the mass outflow rate by
\begin{equation}
    \frac{{\rm d}\dot{M}_{\rm acc}(r)}{{\rm d}r}=4\pi r\dot{m}_{\rm jet}(r).
\label{eq:5}
\end{equation}

These outflow matter will be accelerated magnetically to a Lorentz factor $\Gamma_{\rm BP}$. Following Cao (\cite{Cao02}), the relation between mass flux $\dot{m}_{\rm{jet}}$ and the Lorentz factor of the jet $\Gamma_{\rm BP}$ is
\begin{equation}
    \dot{m}_{\rm{jet}}=\frac{(B_{\rm{D}}^{\rm{P}})^2}{4\pi}(r\Omega_{\rm{D}})^\alpha\frac{\Gamma_{\rm BP}^\alpha}{(\Gamma_{\rm BP}^2-1)^{(1+\alpha)/2}},
\label{eq:6}
\end{equation}
\noindent where $\Gamma_{\rm{BP}}$ is the Lorentz factor of the outer jet. The quantity $\Omega_{\rm{D}}$ is the Keplerian angular velocity at the foot point of the field line:
\begin{equation}
    \Omega_{\rm{D}}=\frac{1}{M({\xi}^{3/2}\chi_{\rm ms}^3+a_{*})},
\label{eq:7}
\end{equation}
\noindent where $\xi\equiv r/r_{\rm{ms}}$ is a radial parameter of the disk defined in terms
of the radius $r_{\rm{ms}}$, and $\chi_{\rm{ms}}$ is defined as $\chi_{\rm{ms}}\equiv \sqrt{r_{\rm{ms}}/M}$.

Substituting equations (\ref{eq:1})-(\ref{eq:6}) into equation (\ref{eq:7}), we obtain the Lorentz factor of the BP jet at disk radius $r$
\begin{equation}
    \frac{\Gamma_{\rm{BP}}^\alpha}{(\Gamma_{\rm{BP}}^2-1)^{(1+\alpha)/2}}=
     \frac{s}{2}\frac{{\xi}^{s+1/2}}{\xi_{\rm H}^{1/2}}\left(\frac{\xi\chi_{\rm ms}^2}{{\xi}^{3/2}\chi_{\rm ms}^3+a_{*}}\right)^{-\alpha}.
\label{eq:8}
\end{equation}

\begin{figure}
  \centering
  %\begin{minipage}{80mm}
  {\includegraphics[width=48mm]{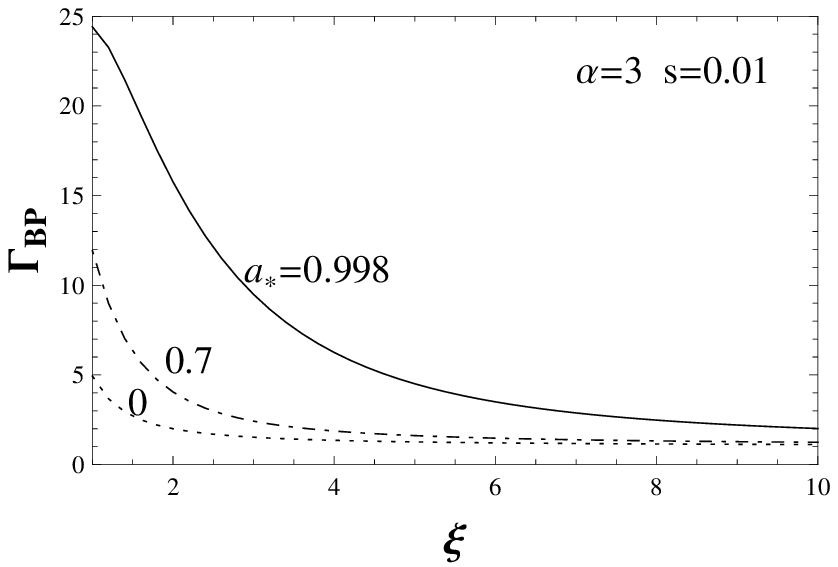}\hfill
  \includegraphics[width=48mm]{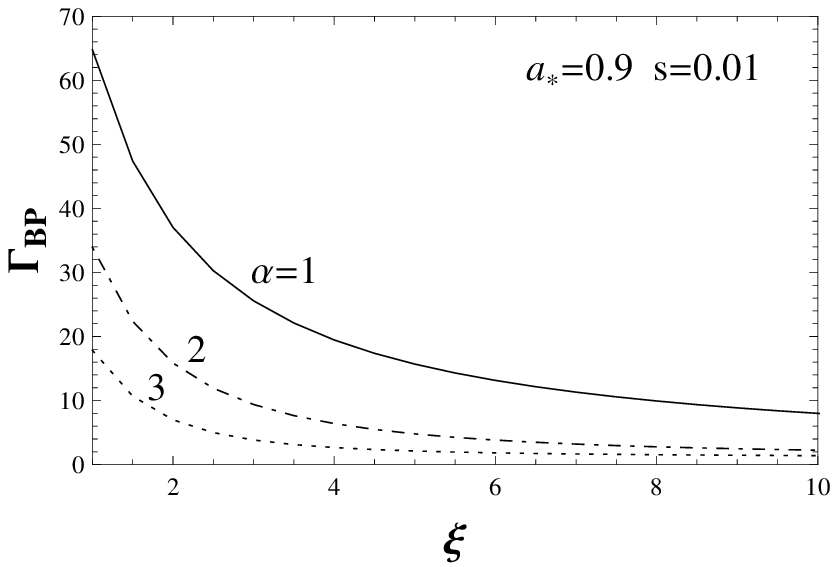}\hfill
  \includegraphics[width=48mm]{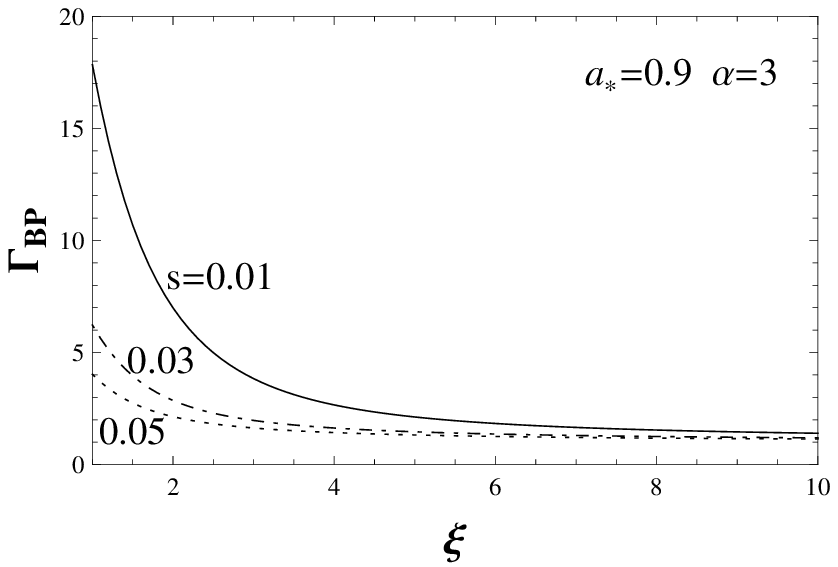}
  \centerline{\hspace{3mm}(a)\hspace{47mm}(b)\hspace{47mm}(c)}
  }
  \caption{The curves $\Gamma_{\rm{BP}}$ versus disk radius $\xi$ for different values of $a_{*}$ (panel a, where $\alpha=3$ and $s=0.01$ are fixed), $\alpha$ (panel b, where $a_*=0.9$ and $s=0.01$ are fixed) and $s$ (panel c, where $a_*=0.9$ and $\alpha=3$ are fixed).}
  %\end{minipage}
\label{Fig:fig2}
\end{figure}

It is shown in equation (\ref{eq:8}) that the distribution of $\Gamma_{\rm BP}$ with disk radius $r$ depends on three parameters: the BH spin $a_{*}$, the self-similar index $\alpha$, and $s$. The curves of $\Gamma_{\rm BP}$ versus $r$ for different values of $a_{*}$,  $\alpha$ and $s$ are shown in Fig.~\ref{Fig:fig2}.

From Fig.~\ref{Fig:fig2} we find that $\Gamma_{\rm BP}$ decreases with the increasing disk radius $r$. This is reasonable, since magnetic acceleration mostly occurs in the inner region. For higher BH spin $a_*$, the inner disk become closer to the BH where the magnetic field is stronger, and the effective acceleration region is expanded consequently (we take the zone between the accretion disk and Alfv\'{e}n surface as the effective acceleration region). In addition, a greater $a_{*}$ indicates a faster Keplerian rotational angular velocity of the disk which results in a larger centrifugal force. We therefore expect larger $\Gamma_{\rm BP}$ for greater $a_*$ (see Fig.~\ref{Fig:fig2}a). Fig.~\ref{Fig:fig2}b shows that $\Gamma_{\rm BP}$ decreases with increasing $\alpha$. This is physically reasonable since a larger $\alpha$ represents a steeper poloidal magnetic field configuration which results in a less efficient acceleration of the disk wind. The third parameter, $s$, is related to the mass loss rate. A larger value of $s$ implies a stronger baryon loading, and this leads to a jet with smaller ${\Gamma_{\rm BP}}$ (as shown in Fig.~\ref{Fig:fig2}c).
\subsection{The Lorentz Factor of the inner jet driven by the BZ process}
%\subsubsection{The BZ power from unit area of the BH horizon surface}
The BZ power transferred through two adjacent magnetic surfaces between $\theta$ and $\theta+{\rm d}\theta$ on the BH horizon is given as (Wang et al. \cite{Wang02}; Lei et al. \cite{Lei07})
\begin{equation}
    {\rm d}P_{\rm{BZ}}=2k(1-k)(B_{\rm H}^{\rm P})^2 M^2a_{*}^2\frac{\sin^3\theta}{2-(1-q)\sin^2\theta}{\rm d}\theta,
\label{eq:9}
\end{equation}
\noindent where $q \equiv \sqrt{1-a_{*}^2}$, and $k\equiv\Omega_{\rm F}/\Omega_{\rm H}$ denotes the ratio of angular velocity of magnetic field line to BH horizon. Usually, we take $k=0.5$ which corresponds to the maximum BZ power. The BZ power from unit area of the horizon is expressed as\\
\begin{equation}
    \widetilde{P}_{\rm{BZ}}=\frac{{\rm d}P_{\rm BZ}}{2{\rm d}S},
\label{eq:10}
\end{equation}
\noindent in which the loop area ${\rm d}S$ is defined by
\begin{equation}
    {\rm d}S=2\pi\tilde{\omega}_{\rm H}\rho_{\rm H}{\rm d}\theta=4{\pi}Mr_{\rm H}\sin\theta{\rm d}\theta.
\label{eq:11}
\end{equation}
\noindent Substituting equations (\ref{eq:9}) and (\ref{eq:11}) into equation (\ref{eq:10}), we have
\begin{equation}
     \widetilde{P}_{\rm{BZ}}=\frac{(B_{\rm H}^{\rm P})^2}{16\pi}\frac{(1-q)\sin^2\theta}{2-(1-q)\sin^2\theta}.
\label{eq:12}
\end{equation}

Due to lack of detailed knowledge of baryon loading and particle acceleration in the BZ process, we take the following assumptions: i) all of the matter entrained into the BZ jet come from the inner edge of the accretion disk, and the mass injection rate is a fraction of the mass accretion rate at ISCO; ii) the magnetic energy is effectively converted into the kinetic energy of baryons in the jet (Zhang \& Yan \cite{ZhangB11}).

Based on assumption i), we have the relation between the mass flux of the BZ jet and the mass accretion rate at the inner edge of disk as follows,
\begin{equation}
    2\int_{0}^{\theta}\dot{m}_{\rm jet}^{\rm BZ}(\theta')2\pi\tilde{\omega}_{\rm H}\rho_{\rm H}{\rm d}\theta'=f(\theta)\dot{M}_{\rm acc}(r_{\rm ms}),
\label{eq:13}
\end{equation}
\noindent where $f(\theta)$ denotes the fraction of accreting mass serving as the matter injection of the jet launched from the BH horizon within the angular range $0-\theta$. Equation (\ref{eq:13}) can be written as
\begin{equation}
    4\pi\tilde{\omega}_{\rm H}\rho_{\rm H}\dot{m}_{\rm jet}^{\rm BZ}(\theta)=\dot{M}_{\rm acc}(r_{\rm ms})\frac{{\rm d}f(\theta)}{{\rm d}\theta}.
\label{eq:14}
\end{equation}

\noindent Considering that the mass injection may reduce as the matter gets inside the BZ jet, we then assume $f\left(\theta\right)$ to be an increasing function of the polar angle $\theta$ as follows,
\begin{equation}
    f(\theta)=\eta(1-\cos\theta)^{n}.
\label{eq:15}
\end{equation}
\noindent From equation (\ref{eq:15}), we have $\eta=f(\pi/2)$, so the parameter $\eta $ is the fraction of the mass injection for the total BZ jet. The parameter $n$ is used to adjust the distribution of mass injection in terms of the polar angle $\theta $.
\noindent Combining equations (\ref{eq:14}), (\ref{eq:15}) and $\tilde{\omega}_{\rm H}\rho_{\rm H}=2Mr_{\rm H}\sin\theta$, we have
\begin{equation}
    \dot{m}_{\rm jet}^{\rm BZ}(\theta)=\frac{\eta\dot{M}_{\rm acc}(r_{\rm ms})}{8\pi M r_{\rm H}}n(1-\cos \theta)^{n-1}.
\label{eq:16}
\end{equation}

\noindent According to assumption ii), the Lorentz factor of BZ jet can be expressed as
\begin{equation}
    \Gamma_{\rm{BZ}}(\theta)=1+\frac{\widetilde{P}_{\rm BZ}(\theta)}{\dot{m}_{\rm jet}^{\rm BZ}(\theta)}.
\label{eq:17}
\end{equation}

\noindent Incorporating equation (\ref{eq:17}) with equations (\ref{eq:3}), (\ref{eq:12}), (\ref{eq:16}) we obtain
\begin{equation}
 \Gamma_{\rm{BZ}}(\theta)=1+\frac{(1-q)\sin^{2}\theta}{\eta n(1-\cos \theta)^{n-1}(1+q)[2-(1-q)\sin^{2}\theta]}.
\label{eq:18}
\end{equation}
\noindent A reasonable distribution of $\Gamma_{\rm{BZ}}$ should be a decreasing
function of the polar angle $\theta$, meanwhile, this function should be finite where $\theta=0$, these two constraints correspond to $n=2$. Then equation (\ref{eq:18}) reduces to
\begin{equation}
 \Gamma_{\rm{BZ}}(\theta, \eta, a_{*})=1+\frac{(1-q)(1+\cos \theta)}{2\eta (1+q)[2-(1-q)\sin^{2}\theta]}.
\label{eq:19}
\end{equation}

\begin{figure}
  \centering
  %\begin{minipage}{80mm}
  {\includegraphics[width=70mm]{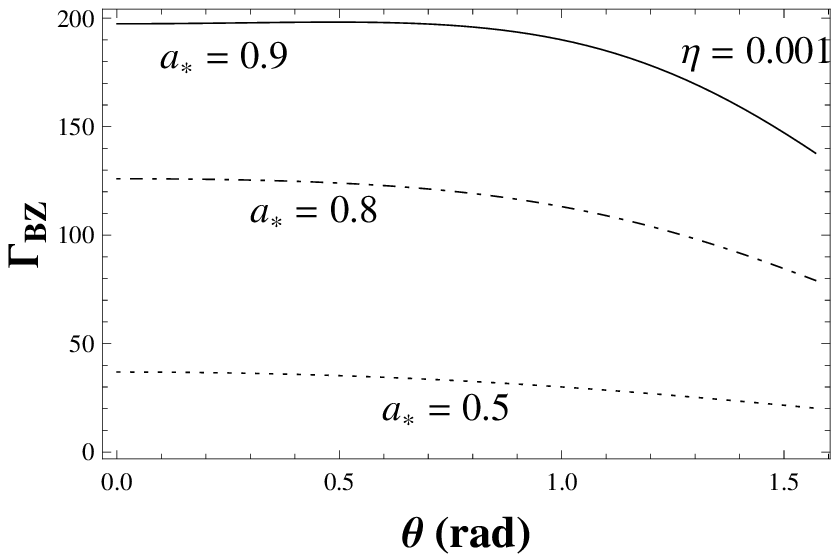}\hfill
   \includegraphics[width=70mm]{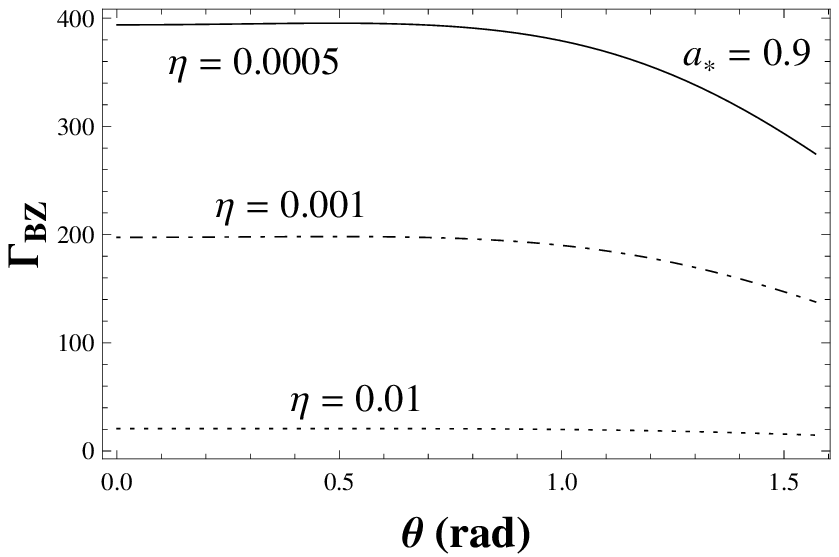}
  \centerline{\hspace{6mm}(a)\hspace{73mm}(b)}
  }\\
  \caption{The curves of $\Gamma _{\rm BZ}$ varying with the polar angle $\theta $ for : (a) different ${{a}_{*}}$, where $\eta=0.001$; (b) different $\eta $, where ${{a}_{*}}=0.9$.}
  %\end{minipage}
\label{Fig:fig3}
\end{figure}

\noindent The curves of ${{\Gamma }_{\rm BZ}}\left( \theta ,{{a}_{*}},\eta  \right)$ varying with the polar angle $\theta $ for different BH spin ${{a}_{*}}$ and efficiency $\eta$ are shown in Fig.~\ref{Fig:fig3}.

The maximum value of the Lorentz factor $\Gamma_{\rm BZ}$ can be obtained by equating the derivative of equation (\ref{eq:19}) to zero. The angle position where the Lorentz factor $\Gamma_{\rm BZ}$ gets its maximum is listed as follows,
\begin{equation}
    \theta_{\rm m}=\left\{\begin{array}{ll}
    0,& 0 \le a_* \le \frac{\sqrt{3}}{2},\\
    \arccos\left(\sqrt{\frac{2}{1-q}}-1\right), & \frac{\sqrt{3}}{2} \le a_* <1,
    \end{array}\right.
\label{eq:20}
\end{equation}
\noindent correspondingly, we have the maximum value of the Lorentz factor $\Gamma_{\rm BZ}$ as
\begin{equation}
    \Gamma_{\rm{BZ}}^{\rm max}=\left\{\begin{array}{ll}
    1+\frac{1-q}{2\eta(1+q)}, & 0 \le a_* \le \frac{\sqrt{3}}{2},\\
    1+\frac{\sqrt{1-q}}{4\eta(1+q)(\sqrt{2}-\sqrt{1-q})}, & \frac{\sqrt{3}}{2} \le a_* <1.
    \end{array}
    \right.
\label{eq:21}
\end{equation}

\begin{figure}
  \centering
  \includegraphics[width=80mm]{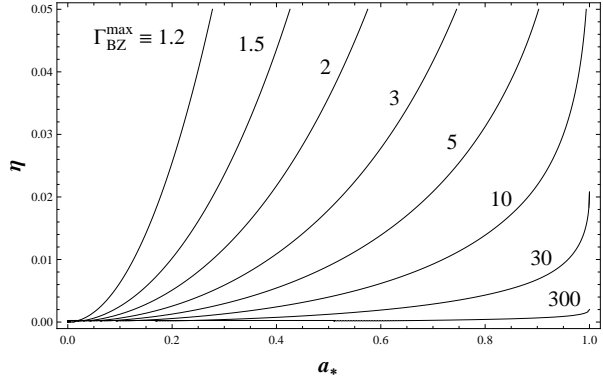}\\
  \caption{The contours of $\Gamma_{\rm BZ}^{\rm max}$ in the parameter space $\left( a_{*}, \eta \right)$.}
\label{Fig:fig4}
\end{figure}

Generally, ${{\Gamma }_{\rm BZ}}$ decreases with the increasing polar angle $\theta $ (as shown in Fig.~\ref{Fig:fig3}), which is consistent with the observations of the structured jets. Fig.~\ref{Fig:fig3}a and Fig.~\ref{Fig:fig3}b show that ${{\Gamma }_{\rm BZ}}$ increases with the increasing BH spin $a_*$, while it decreases with the parameter $\eta$. To make this more clear, we also plot the contours of $\Gamma _{\rm BZ}^{\rm max }$ in the parameter space $\left( {{a}_{*}},\eta  \right)$ as shown in Fig.~\ref{Fig:fig4}.

According to equations (\ref{eq:9}), (\ref{eq:13}) and (\ref{eq:15}), a larger $a_{*}$ implies a stronger BZ power whereas a larger $\eta$ denotes a stronger matter injection into the jet, therefore the above results are physically sensible.

\section{Fitting the Lorentz factors of GRBs and AGNs}
\label{fitting}
\begin{figure}
  \begin{center}
  %\begin{minipage}{50mm}
  {\includegraphics[width=48mm]{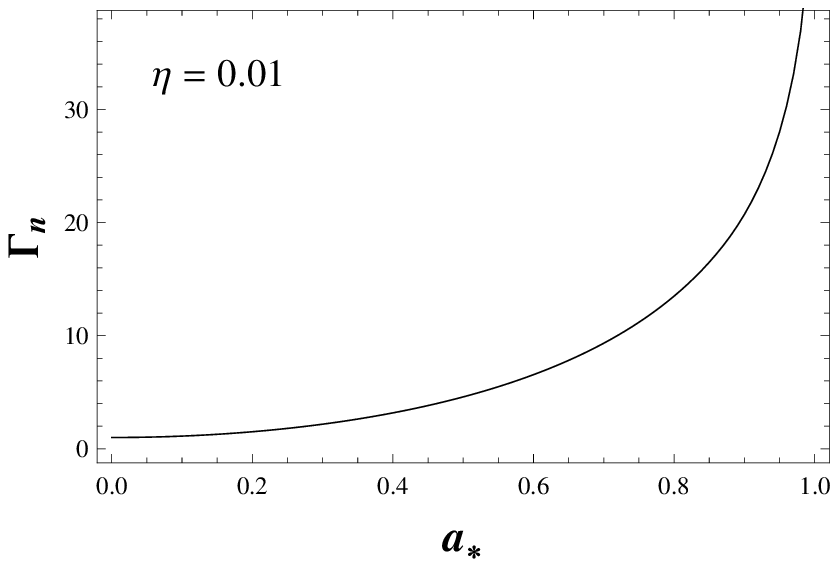}\hfill
  \includegraphics[width=48mm]{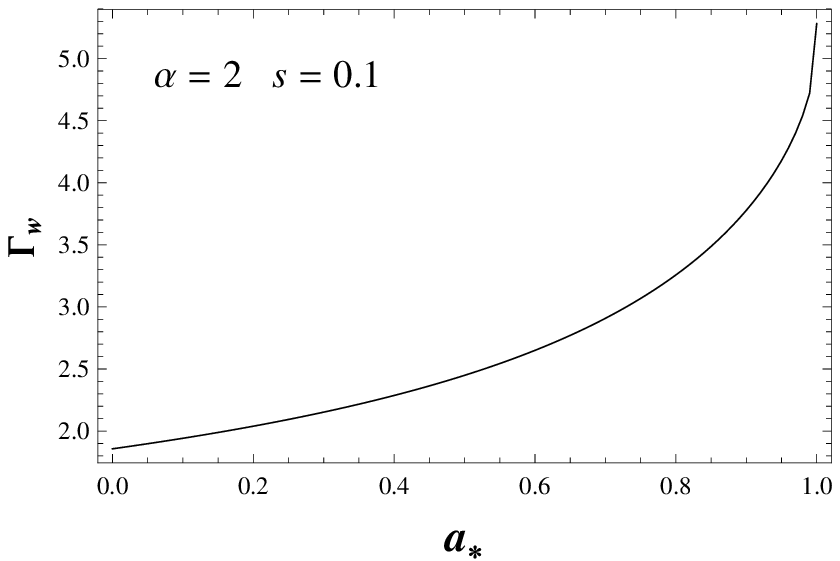}\hfill
  \includegraphics[width=48mm]{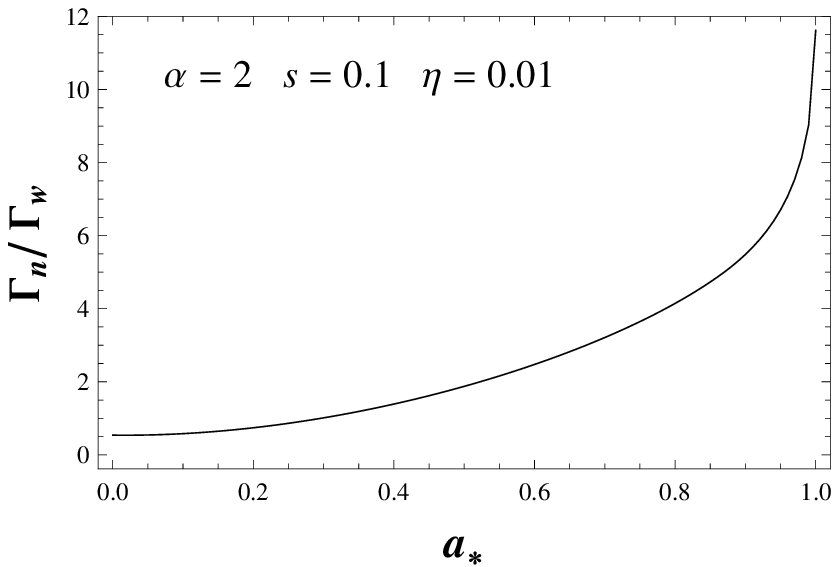}
  \centerline{\hspace{5mm}(a)}
  \includegraphics[width=48mm]{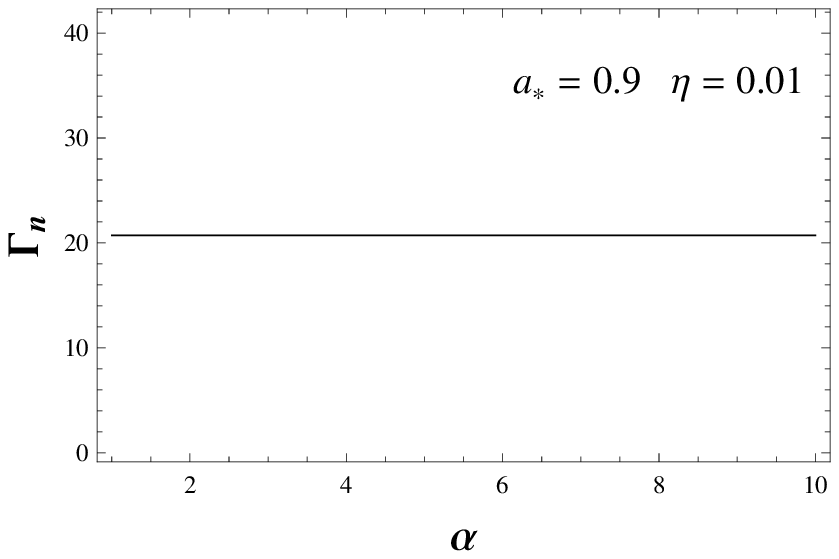}\hfill
  \includegraphics[width=48mm]{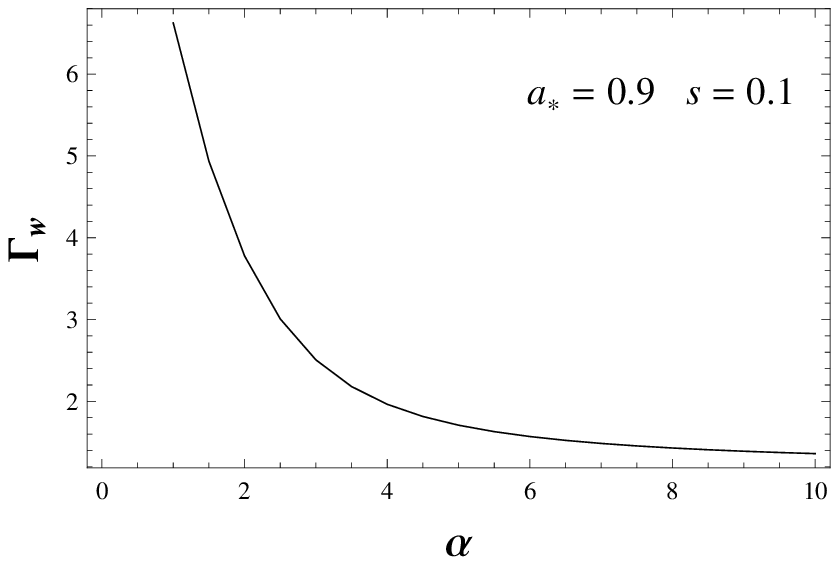}\hfill
  \includegraphics[width=48mm]{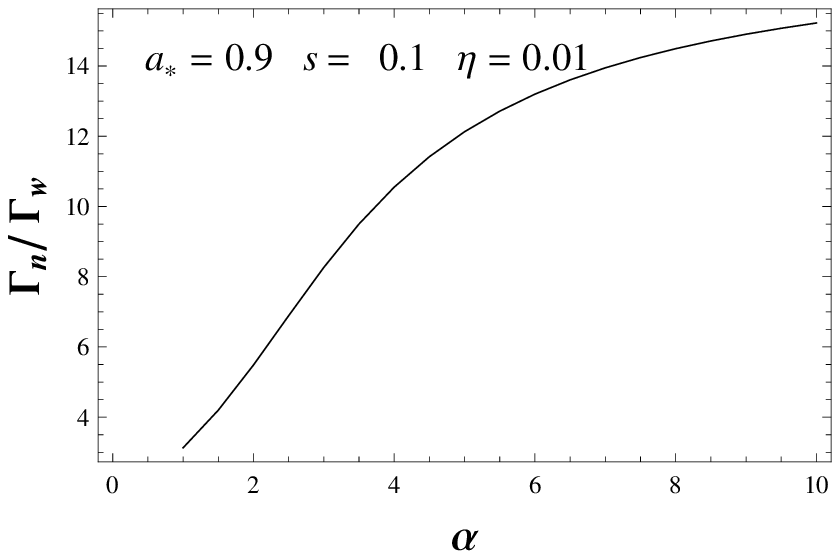}
  \centerline{\hspace{5mm}(b)}
  \includegraphics[width=48mm]{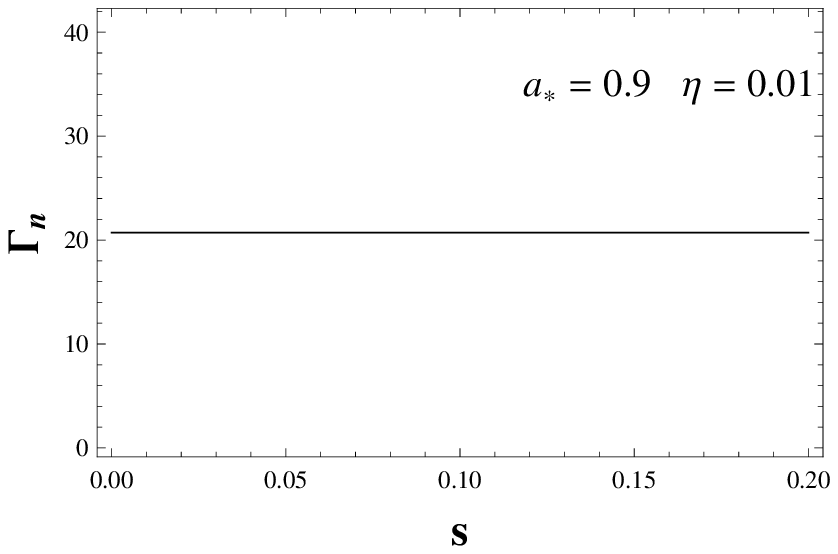}\hfill
  \includegraphics[width=48mm]{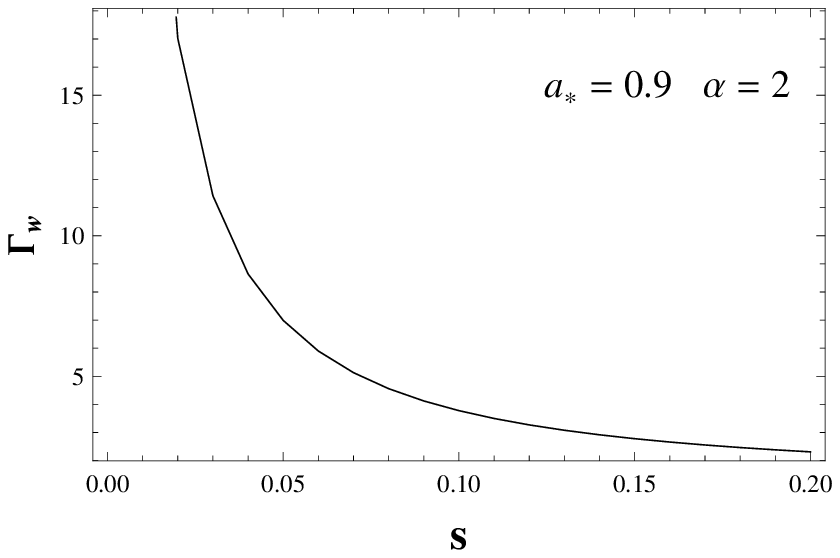}\hfill
  \includegraphics[width=48mm]{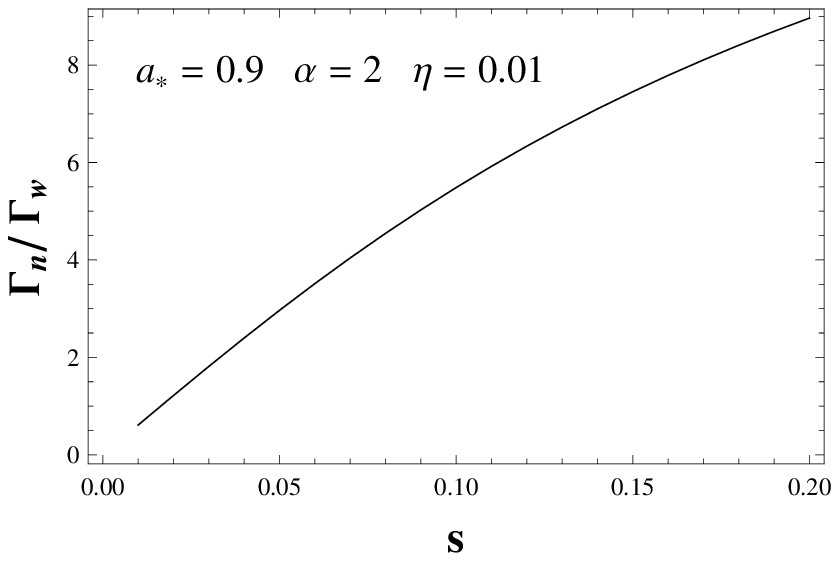}
  \centerline{\hspace{5mm}(c)}
  \includegraphics[width=48mm]{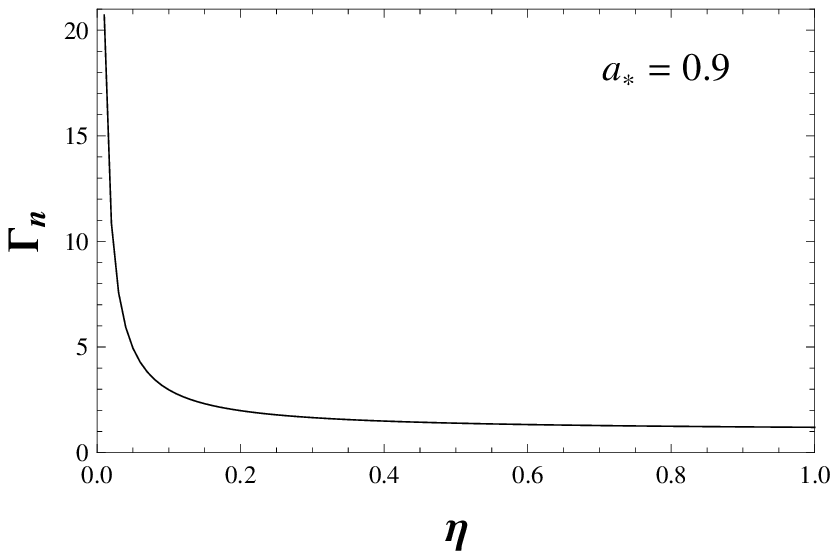}\hfill
  \includegraphics[width=48mm]{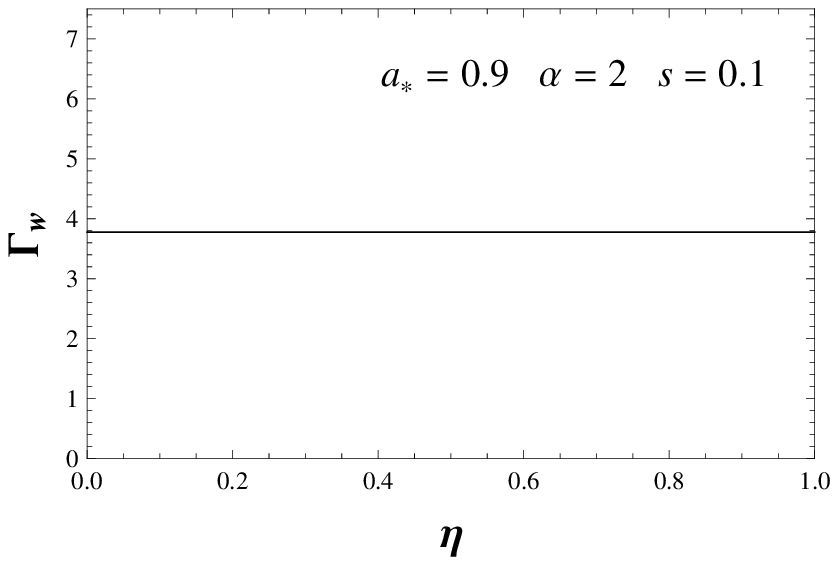}\hfill
  \includegraphics[width=48mm]{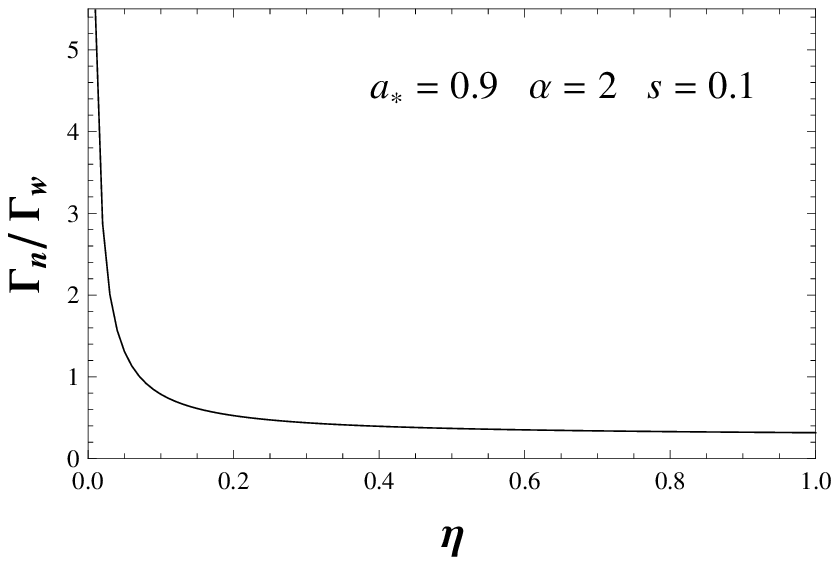}
  \centerline{\hspace{5mm}(d)}
  }
  \caption{The curves of Lorentz factors $\Gamma_n$, $\Gamma_w$ and the ratio $\Gamma_n/\Gamma_w$ versus $a_*$ (panel a, where $\alpha=2$, $s=0.1$ and $\eta=0.01$), $\alpha$ (panel b, where $a_*=0.9$, $s=0.1$ and $\eta=0.01$), $s$ (panel c, where $a_*=0.9$, $\alpha=2$ and $\eta=0.01$) and $\eta$ (panel d, where $a_*=0.9$, $\alpha=2$ and $s=0.1$).}
  %\end{minipage}
  \end{center}
\label{Fig:fig5}
\end{figure}

Inspecting Figs.~\ref{Fig:fig3}a and \ref{Fig:fig3}b, we find the variation of ${{\Gamma }_{\rm BZ}}$ with the angle $\theta $ is very smooth. Therefore, for simplicity, we use $\Gamma_{\rm{BZ}}^{\rm max}$ as the typical value of the Lorentz factor of the narrow fast jet, and $\Gamma_{\rm{BP}}^{\rm max}$ as the typical value of the Lorentz factor of the wide slow jet, and we define $\Gamma_{\rm{n}} \equiv \Gamma_{\rm{BZ}}^{\rm max}$, and $\Gamma_{\rm{w}} \equiv \Gamma_{\rm{BP}}^{\rm max}$. The ratios of the Lorentz factor of narrow fast jet $\Gamma_{\rm{n}}$ to wide slow jet $\Gamma_{\rm{w}}$ are shown in Fig.~\ref{Fig:fig5}.

Fig.~\ref{Fig:fig5}a shows that $\Gamma_{\rm n}$ is greater than $\Gamma_{\rm w}$ for large BH spin $a_*$. Since $a_{*}=0.9$ may be a typical BH spin in object with strong relativistic jet (e.g., Wu et al. \cite{WuQW11}; van Putten \cite{vanPutten04}), we just take $a_*=0.9$ in the following calculations. The ratio $\Gamma_{\rm n}/\Gamma_{\rm w}$ increases with increasing $\alpha$ and $s$, while deceases with increasing $\eta$. These results can be well understood by inspecting the left and middle panels of Figs.~\ref{Fig:fig5}b-d (also see the discussions in Section 2). Therefore, to make a two-component jet with inner-faster and outer-slower structure, the values of $\alpha$ and $s$ should not be too small, and the value of $\eta$ should not be too large. In this paper, we take $\alpha=2$ in calculations, and study the parameters $s$ and $\eta$ for different sources.

Filgas et al. (\cite{Filgas11}) fitted the broad-band light curve of the afterglow of GRB 080413B with an on-axis two-component jet model, and the two components have opening angles of $\theta_{\rm{n}}\sim1.7^\circ$ and $\theta_{\rm{w}}\sim9^\circ$, and Lorentz factors of $\Gamma_{\rm{n}}>188$ and $\Gamma_{\rm{w}}\sim 18.5$. By using our model, we find that $\eta<0.01$ and $s \sim 0.023$. We also study two other GRBs, i.e., GRB 030329, GRB 051221A, and several AGNs (CenA, Mkn 501, and Mkn 421), for which the required Lorentz factor to fit the observations are known. The estimated value for the two parameters $\eta$ and $s$ are listed in Table 1.

%-----------------------------------Table 1----------------------------------------
\begin{table}[htbp]

\caption[]{Fitting the Lorentz factors of the two-component jets from GRBs and AGNs}
\label{Tab:table1}
\begin{minipage}{180mm}
\begin{center}

%\centering
\begin{tabular}{lccccc}
  \hline\noalign{\smallskip}\hline\noalign{\smallskip}
  Source & $\Gamma_{\rm{n}}$ & $\Gamma_{\rm{w}}$ & $\Gamma_{\rm{n}}/\Gamma_{\rm{w}}$ & $\eta$ & $s$ \\
  \hline\noalign{\smallskip}
  GRB 080413B & $>$188 & 18.5 & 10.16 &  $<$0.001 & 0.023 \\
  GRB 030329 & 300 & 30 & 10 &  0.0007 & 0.014 \\
  GRB 051221A & 500 & 50 & 10 &  0.0004 & 0.009 \\
  \hline\noalign{\smallskip}
  CenA & 15 & 3 & 5 &  0.014 & 0.17 \\
  Mkn 501 & 15 & 3.5 & 4.29 &  0.014 & 0.14 \\
  Mkn 421 & 17 & 3 & 5.67 &  0.012 & 0.17 \\
 \noalign{\smallskip}\hline
  %\multicolumn{9}{1}{}

\end{tabular}\\

Notes: the Lorentz factors of the two-component jets of the above sources are quoted from the following references: Filgas et al. (\cite{Filgas11}) for GRB 080413B, Huang et al. (\cite{Huang06}) for GRB 030329, Jin et al. (\cite{Jin07}) for GRB 051221A, Ghisellini et al. (\cite{Ghisellini05}) for Cen A, Mkn 501 and Mkn 421. In our calculations, we take $n=2, a_{*}=0.9$ and $\alpha=2$.
\end{center}
\end{minipage}
\end{table}

From Table~\ref{Tab:table1}, we find that the values of s and �� for GRBs are much smaller than those for AGNs. For GRBs, the typical value of $\eta$ is about 0.0001, and that of $s$ is about 0.01, while for AGNs the above two values become $\eta \sim 0.01$ and $s \sim 0.15$. The big difference arises from the Lorentz factors of GRBs are much greater than those of AGNs.

\section{Conclusion and Discussion}
\label{sect:conclusion and discussion}
In this paper, we propose a two-component jet model by combining the BZ and BP processes. We find that the Lorentz factor of the jet driven by the BZ process possess is generally greater than that of the jet driven by the BP process. Therefore, our model provides a natural explanation for the origin of the inner-narrow-fast and the outer-wide-slow jets. We then fit the Lorentz factors of several GRBs and AGNs, which are believed to be powered by two-component jet. It turns out that the physical parameters related to the central engine of these objects can be constrained to a narrow range.

For GRBs, the typical value of $\eta$ is about 0.0001, and that of the parameter $s$ is about 0.01, while for AGNs the above two typical values become $\eta \sim 0.01$ and $s \sim 0.15$.

The values of $s$ for GRBs are much smaller than those for AGNs. This result can be understood as follows. Although the jet physics for GRBs and AGNs may be similar, the accretion modes for the two types of sources are however rather different. For AGN sources, the accretion rates are less than Eddington accretion rate, and the accretion mode is probably advection dominated by fitting the luminosity and spectral features. The property of an advection dominated disk is that it has a strong wind which is driven by a positive Bernouilli constant (Narayan \& Yi \cite{Narayan94}). To model the radiatively inefficient accretion flow in the Galactic Center source Sgr A*, Yuan, Quataert \& Narayan (\cite{Yuan03}) deduced $s\sim0.3$, which is close to our results for AGN sources. However, GRBs always involve a hyperaccreting disk, which is dominated by neutrino cooling rather than advection (Popham, Woosley \& Fryer \cite{Popham99}). This kind of disk can only drive a weak wind by neutrino heating or magnetic centrifugal force.

Although this model provides a clear picture for the two-component jet, it is excessively simplified in the following aspects. Firstly, the further acceleration after the Alfv\'{e}n point is ignored in the calculation for the BP process, but in reality there might be other acceleration procedures, e.g., magnetic pressure gradient. Secondly, the details of acceleration is not taken into account in fitting the narrow jet to avoid the complicated MHD calculation. Instead, we assume that all matter injecting into the narrow jet comes from the region within ISCO, and most of the electromagnetic energy is converted into the kinetic energy of the jet matter. Thirdly, we do not give the opening angles of the two-component jet separately, which are very important parameters in fitting the light curves of the afterglows of GRBs. Fourthly, we do not discuss the interaction between the inner and outer jets, which may influence the high energy radiation spectra of AGNs. Disk accretion dynamics and numerical simulation is needed for more sophisticated solution.

\begin{acknowledgements}
    We are very grateful to an anonymous referee for many useful comments and suggestions, which allowed us to improve the manuscript substantially. This work is supported by National Natural Science Foundation of China (grants 11173011, 11003004, 10703002, 11143001, 11133005 and 11103003), National Basic Research Program (``973'' Program) of China grant 2009CB824800, and Fundamental Research Funds for the Central Universities (HUST: 2011TS159). WHL acknowledges a Fellowship from China Scholarship Program for support.
\end{acknowledgements}

%\subsection{Please Capitalize the First Letter of Each Notional Word in Subsection Title}
%\subsubsection{This is a third-level section --- subsubsection}
%Some applications of the routines are given in Table~\ref{Tab:publ-works}.

%\appendix                  %%appendicial material is supported

%\section{This shows the use of appendix}
%A postscript file is actually an ASCII text file (you may even edit it).
%However, you need to transfer a PDF file or any compressed or packaged
%file in binary mode when using FTP.

%\section{What is SCI?}
%SCI is the abbreviation of Science Citation Index system powered by
%the Institute for Scientific Information (ISI). For details please
%visit {\it http://apps.isiknowledge.com}.

\label{lastpage}

\end{document}